\documentclass{aa}  

\usepackage{graphicx}
\usepackage{multirow}
\usepackage{caption}
\usepackage{multirow}
\usepackage{natbib}
\usepackage{longtable}
\usepackage{natbib}

\newcommand{\teff}{$T_{\rm eff}$}

\newcommand{\tef}{T_\mathrm{eff}}

\newcommand{\feh}{\mathrm{[Fe/H]}}

\newcommand{\qq}{$\mathrm{q}^{2}$}
\newcommand{\tm}{$\rm{M_{\oplus}}$}
\newcommand{\sm}{$\rm{M_{\odot}}$}
\newcommand{\vsini}{$v \sin i$}

\bibpunct{(}{)}{;}{a}{}{,}

\graphicspath{{plots/}}
\usepackage{txfonts}
\begin{document} 

 \title{High-precision analysis of the solar twin HIP 100963}

   \author{Jhon Yana Galarza
          \inst{\ref{inst1}}
          \and
          Jorge Mel\'endez\inst{\ref{inst1}}
          \and
          Ivan Ram\'{i}rez\inst{\ref{inst2}}
          \and
          David Yong \inst{\ref{inst3}}
          \and 
          Amanda I. Karakas \inst{\ref{inst3}}
          \and
          Martin Asplund \inst{\ref{inst3}}
          \and 
          Fan Liu \inst{\ref{inst3}} }
          
 \institute{Universidade de S\~ao Paulo, IAG, Departamento de Astronomia, Rua do Mat\~ao 1226, S\~ao Paulo, 05509-900 SP, Brasil \\ \email{ramstojh@usp.br}\label{inst1}
 \and University of Texas at Austin, McDonald Observatory and Department of Astronomy, 2515 Speedway, Austin, TX 78712-1205, USA \label{inst2}
 \and Australian National University, Research School of Astronomy and Astrophysics, Mt. Stromlo Observatory, via Cotter Rd., Weston, ACT 2611, Australia \label{inst3}
}

\titlerunning{High precision analysis of the solar twin HIP 100963}

%\date{Received September 15, 1996; accepted March 16, 1997}

  \abstract
   {HIP100963 was one of the first solar twins identified. Although some high-precision analyses are available, a comprehensive high-precision study of chemical elements from different nucleosynthetic sources is still lacking from which to obtain potential new insights on planets, stellar evolution, and Galactic chemical evolution (GCE).}
   {We analyze and investigate the origin of the abundance pattern of HIP 100963 in detail, in particular the pattern of the light element Li, the volatile and refractory elements, and heavy elements from the $s$- and $r$-processes.}
   {We used the HIRES spectrograph on the Keck I telescope to acquire high-resolution (R $\approx$ 70000) spectra with a high
signal-to-noise ratio (S/N $\approx$ 400 - 650 per pixel) of HIP 100963 and the Sun for a differential abundance analysis. We measured the equivalent widths (EWs) of iron lines to determine the stellar parameters by employing the differential spectroscopic equilibrium. We determined the composition of volatile, refractory, and neutron-capture elements through a differential abundance analysis
with respect to the Sun.}  %The lithium abundance was obtained by spectral synthesis.}
   {The stellar parameters we found are $T_{\rm{eff}}=5818 \pm 4$ K, log $g = 4.49 \pm 0.01$ dex, $v_{t} = 1.03 \pm 0.01 $ $\rm{km\ {s}}^{-1}$ , and [Fe/H] $ = -\ 0.003 \pm 0.004$ dex. These low errors allow us to compute a precise mass ($1.03^{+0.02}_{-0.01}$ \sm) and age (2.0 $\pm$ 0.4 Gyr), obtained using Yonsei-Yale isochrones. Using our [Y/Mg] ratio, we have determined an age of $2.1 \pm 0.4$ Gyr, in agreement with the age computed using isochrones. Our isochronal age also agrees with the age determined from stellar activity (2.4 $\pm$ 0.3 Gyr). We study the abundance pattern with condensation temperature ($\rm{T_{cond}}$) taking corrections by the GCE into account. We show that the enhancements of neutron-capture elements are explained by contributions from both the $s$- and $r$-process. The lithium abundance follows the tight Li-age correlation seen in other solar twins.}
   {We confirm that HIP 100963 is a solar twin and demonstrate that its abundance pattern is about solar after corrections for
GCE. The star also shows enrichment in $s-$ and $r$-process elements, as well as depletion in lithium that is caused by stellar evolution.} 


   \maketitle
%
%________________________________________________________________
\section{Introduction}
\label{sec:1}
\citet{Cayrel:1996} defined a solar twin as a star very similar to the Sun within the observational errors, in different properties such as mass, chemical composition, age, effective temperature, luminosity, surface gravity, chromospheric activity, equatorial rotation, etc. The first four stars reported as solar twins were 18 Sco \citep{Porto:1997}, HIP78399 \citep{King:2005}, HD 98618 \citep{Melendez:2006}, and HIP 100963 \citep{Takeda:2007}. \citet{Ramirez:2009} suggested a new definition: a star is a solar twin if its stellar parameters fall into the range of $\Delta T_{\rm{eff}} = 100$ K, $\Delta$log $g = 0.1$ dex, and $\Delta$[Fe/H] = 0.1 dex, relative to the Sun. With this new constraint as definition, there are about 100 solar twins known up to now \citep{Melendez:2007, Melendez:2012, Pasquini:2008, Petit:2008, Melendez:2009, Ramirez:2009, Takeda:2009, Baumann:2010, Onehag:2010, Datson:2012, Nascimento:2013, MelendezSchirbel:2014, Porto:2014, Ramirez:2014-1, Mahdi:2016}.

HIP 100963 was one of the 118 solar analogs\footnote{Solar analogs are stars with temperatures and metallicities  within $\sim$ 500 K and 0.3 dex of the Sun.} observed by \citet{Takeda:2007}. The authors determined its stellar parameters, rotational velocity, Li abundance, luminosity, mass, and age. These last two parameters were computed by comparing the position of the luminosity (L) and effective temperature (\teff) on the theoretical Hertzsprung–Russell diagram. Their first results suggested that HIP 100963 has an age of $\sim$ 5 Gyr with similar stellar parameters as the Sun. Later, \citet{Takeda:2009} obtained the stellar parameters for HIP 100963 with greater precision. Their results show somewhat higher effective temperature ($+40$ K) and higher Li abundance ($+0.7$ dex) than the Sun. The authors also analyzed the Be II lines (3130.42 $\AA$ and 3131.07 $\AA$) and concluded that the Be abundance of HIP 100963 is presumably solar within 0.1 dex\footnote{No Be abundances were given, only a comparison of observed and synthetic spectra was made.}. \citet{Nascimento:2009} determined the age and mass of HIP 100963 employing the Toulouse-Geneva stellar evolution code \citep{Hui-Bon-Hoa:2008}. They used the stellar parameters found by \citet{Takeda:2007} and \citet{Takeda:2009}. Their results agree with the mass derived by \citet{Takeda:2007}, but the age is 3 Gyr, which is younger than the value found by \citet{Takeda:2007} using less precise stellar parameters. Using chromospheric activity, an even younger age (1.7 Gyr) has been suggested \citep{Isaacson:2010}. 
\begin{figure*}
\centering
\includegraphics[width=17cm]{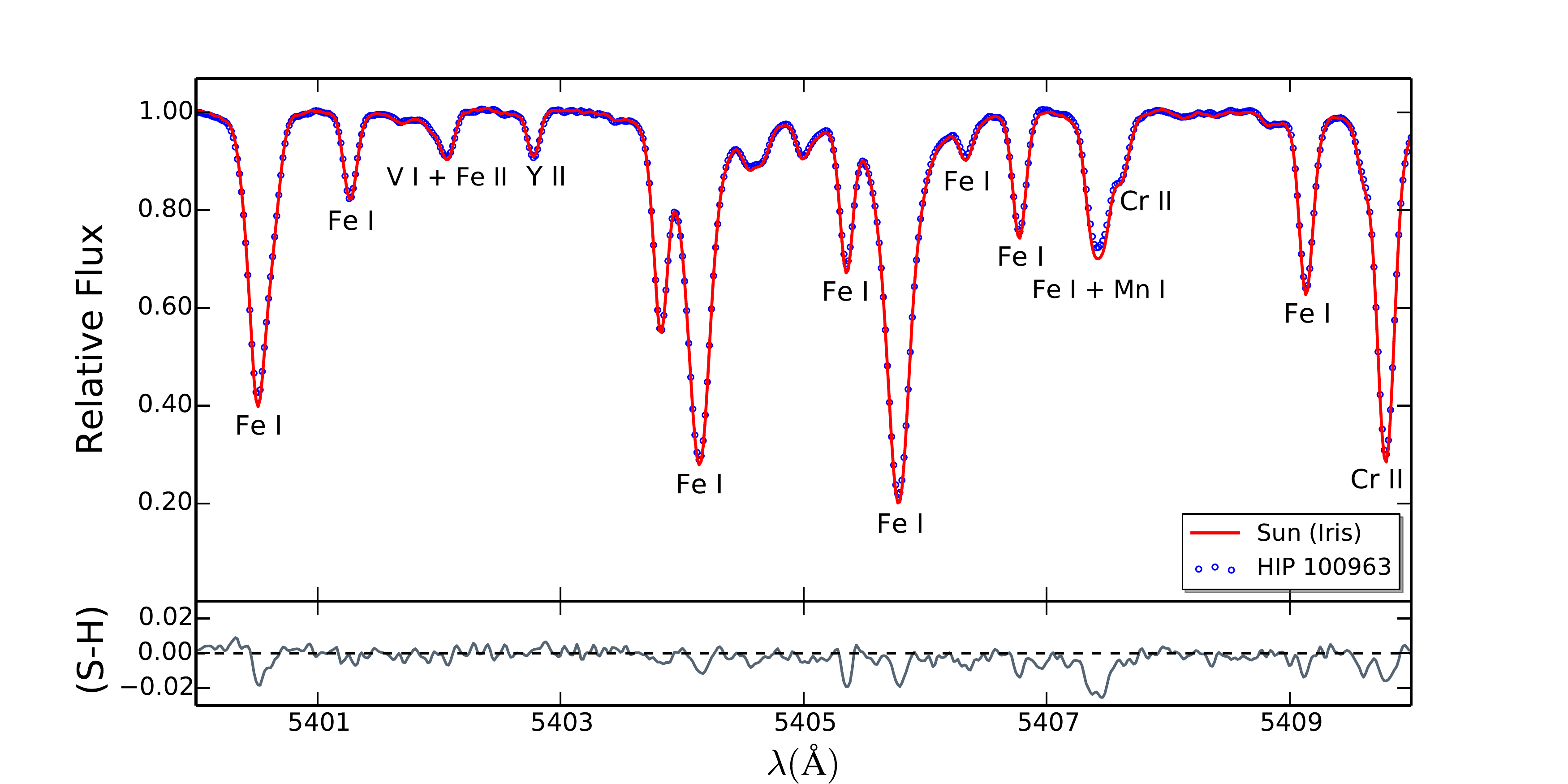} %\\
\caption{HIRES spectra of HIP 100963 (blue circles) and the Sun (red line) in the 5400 - 5410 $\AA$ region. Residuals between the Sun and HIP 100963 (Sun - HIP) are shown in the lower panel.}
\label{fig:compare}
\end{figure*}

In recent works based on spectroscopy of solar-type superflare stars, \citet{Notsu-1:2015} carried out high-dispersion spectroscopic observations for a sample of 50 solar-type stars and 8 solar twins (among them HIP 100963). They determined the stellar parameters of HIP 100963 (see Table \ref{tab:sources}) following the method by \citet{Takeda:2005}. The rotational velocity, chromospheric activity \citep{Notsu-2:2015}, and Li abundance \citep{Notsu-3:2015} were also studied.

\citet{Ramirez:2009} analyzed the spectra of 64 solar-type stars (HIP 100963 among the sample), finding a correlation between the differential abundances relative to the Sun and the condensation temperature. \citet{Melendez:2009} suggested that this deficiency in refractory elements could be explained by rocky planet formation. \citet{Chambers:2010} has reproduced such abundance behavior by adding a mixture of 4 \tm \ of Earth-like material and carbonaceous-chondrite-like material into the convection zone of the Sun. In addition to the planet signatures, the Galactic chemical evolution (GCE) can have a strong impact on the abundance pattern \citep[e.g.,][]{Adibekyan:2014A, Nissen:2015, Spina:2015}, as most elements show [X/Fe] ratios that depend on age. Therefore, it is important to continue to identify more solar twins and to conduct high-precision chemical abundance analyses to better understand the relative contributions of the GCE and planetary signatures to be able to distinguish these effects.

In the present work, we determine the stellar parameters of HIP 100963 with high precision to obtain its mass and age, and clarify its evolutionary status in this way. We also derive precise chemical abundances of 27 elements and investigate how the possible presence of planets, stellar evolution, and the GCE can explain the abundance pattern of this solar twin.
%__________________________________________________________________

\section{Observations and data reduction}
\label{sec:2}
Spectra of HIP 100963 and the Sun (reflected light from the Iris asteroid) were obtained using the HIRES spectrograph \citep{Vogt:1994} at the Keck I telescope, covering the wavelength region from 3940-8350 $\AA$, in a mosaic of three CCDs optimized for the blue, green, and red regions. Observations were performed on August 13, 2013, with exposure times of 180 s for HIP 100963 and 720 s for Iris. The slit width was set to 0.57 arcsec, giving a spectral resolving power of R = $\lambda / \Delta \lambda$ = 70000 (3.2 pixels per resolution element); the signal-to-noise ratio (S/N) at $\sim$ 6000 \AA\ measured for HIP 100963 is about 400 $\rm{pixel}^{-1}$ , while for the Sun it is about 650 $\rm{pixel}^{-1}$.

The spectra were reduced using the MAKEE\footnote{MAKEE was developed by T. Barlow to reduce Keck HIRES spectra. It is accessible at \object{www.astro.caltech.edu/$\sim$tb/}} pipeline following the standard procedure: bias subtraction, flat fielding, sky subtraction, order extraction, and wavelength calibration. MAKEE performs the heliocentric correction for wavelength. Radial velocity and the rest frame correction were performed using the {\tt rvidlines} and {\tt dopcor} tasks in IRAF. The continuum-normalization was performed using the {\tt continuum} task in IRAF\footnote{IRAF is distributed by the National Optical Astronomy Observatory, which is operated by the Association of the Universities for Research in Astronomy, Inc. (AURA) under cooperative agreement with the National Science Foundation.} . Part of the reduced spectra of HIP 100963 and the Iris asteroid is shown in Fig. \ref{fig:compare}.
%__________________________________________________________________
\section{Abundance analysis}
\label{sec:3}
\subsection{Stellar parameters}
The analysis is based on the line-by-line differential method \citep[e.g.,][]{Melendez:2012, Monroe:2013, Liu:2014, Ramirez:2014-1, MelendezIvanAmanda:2014, Biazzo:2015, Nissen:2015, Spina:2015, Saffe:2015}. We measured the EWs using the {\tt splot} task in IRAF, fitting the line profiles using Gaussians. Pseudo-continuum regions were obtained following \cite{Bedell:2014} in a window of 6 $\AA$. 
\begin{table*}
\caption{Comparison of stellar parameters of HIP 100963}
\label{tab:sources}
\centering 
\renewcommand{\footnoterule}{}  % to avoid a line before footnotes
\begin{tabular}{lrrrrrrrl} 
\hline    
\hline 
 $T_{\rm{eff}}$  & error   &  log $g$ & error & $\rm{[Fe/H]}$& error & $v_{t}$ & error &  Source  \\
    (K)     &   (K)   &   (dex)  & (dex) & (dex)        & (dex) &  (km $\rm{s}^{-1}$)       &   (km\ $\rm{s}^{-1}$)           \\
\hline
    5818    &   4     &  4.49  &  0.01       &  -0.003       & 0.004       &  1.03   & 0.01        &  This work  \\
    5775    & 17 - 51 &  4.41  & 0.04 - 0.12 &  -0.012       & 0.02 - 0.06 &  0.98   & 0.09 - 0.27 & \citet{Takeda:2007}*  \\
    5815    &   10    &  4.46  &  0.01       &  -0.010       & 0.005       &  1.01   & 0.04        & \citet{Takeda:2009}$\dagger$   \\ 
    5815    &   50    &  4.49  &  0.07       &   0.018       & 0.019       & \ldots  & \ldots      & \citet{Ramirez:2009}   \\ 
    5834    &   25    &  4.56  &  0.06       &   0.010       & 0.020       &  1.07   & 0.09        & \citet{Notsu-1:2015} \\
\hline       
\hline
\end{tabular}
\tablefoot{\\ * Individual errors were not reported by \citet{Takeda:2007}; the errors vary in the range shown. \\ $\dagger$ The parameters reported here are based on the differential stellar parameters and adopting for the Sun \teff = 5777 K, log $g = 4.44$ dex, and $v_{t} = 1.00\ $km$\ \rm{s}^{-1}$.}
\end{table*}

\begin{figure}
\centering
\includegraphics[width=9cm]{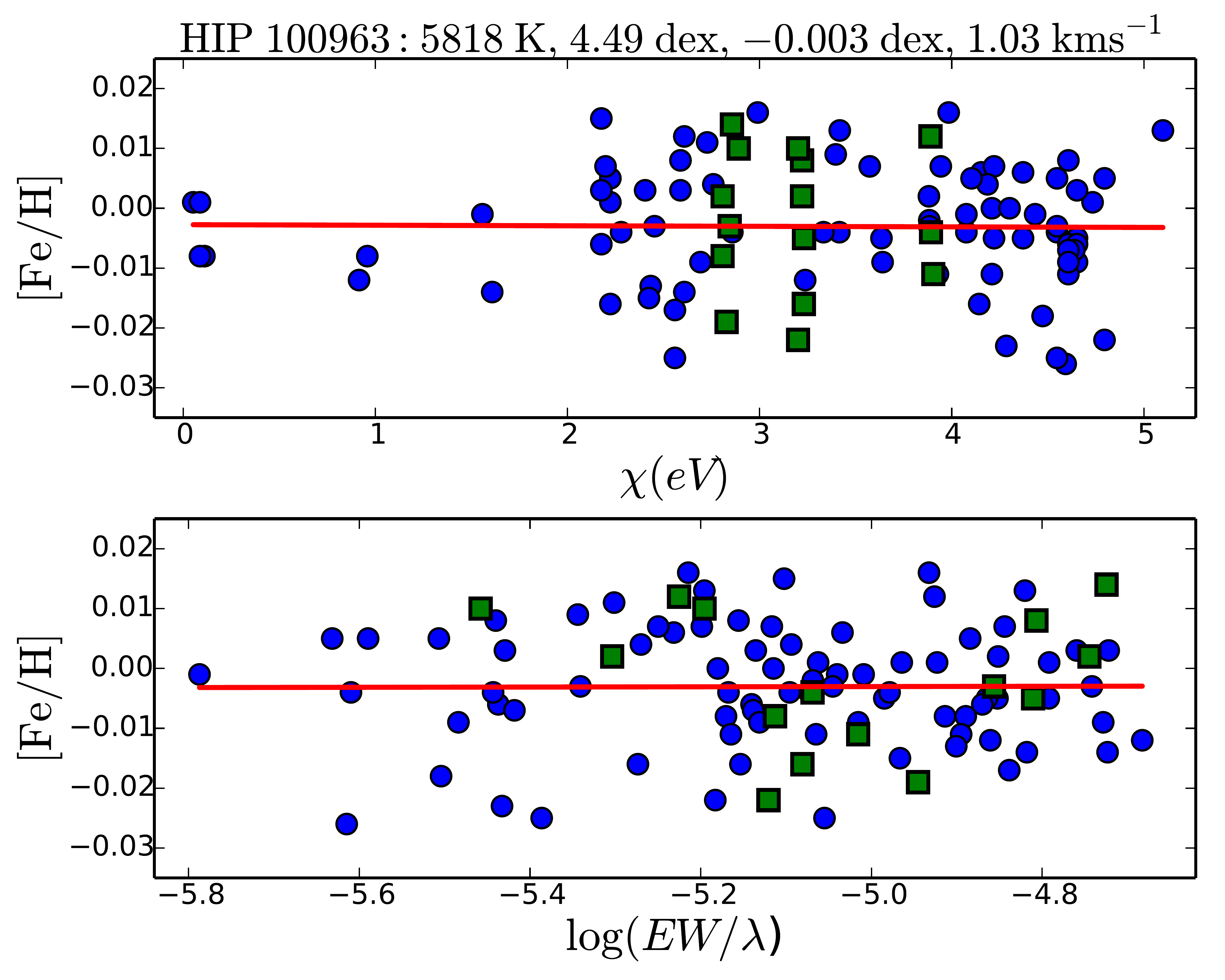}
\caption{Differential abundance of HIP 100963 relative to the Sun as a function of excitation potential (top panel) and reduced equivalent width (bottom panel). The blue filled circles represent Fe I, the green filled squares Fe II. The red solid lines in both panels are linear fits to Fe I. }
\label{fig:parameters}
\end{figure}

To calculate the stellar parameters and chemical abundances, we used the Kurucz ODFNEW model atmospheres \citep{Castelli:2004} and the 2014 version of the local thermodynamic equilibrium (LTE) code MOOG \citep{Sneden:1973}. We determined the elemental abundances of the Sun assuming the standard solar parameters: \teff = 5777 K, log $g = 4.44$ dex \citep{Cox:2000} and $v_{t} = 1.00\ $km$\ \rm{s}^{-1}$, as in \citet{Ramirez:2014-1}. We employed the differential spectroscopic equilibrium\footnote{Spectroscopic equilibrium is achieved when the differential excitation, ionization balance, and  no dependence of differential iron abundance with reduced equivalent width (EW/$\lambda$) are reached.} for HIP 100963 relative to the Sun (see Fig. \ref{fig:parameters}), giving as a result the following stellar parameters: \teff = 5818 K, log $g = 4.49$ dex, $v_{t} = 1.03 $ km$\ \rm{s}^{-1}$ , and [Fe/H] $ = -0.003$ dex. The stellar parameter errors were determined as in \citet{Epstein:2010} and \citet{Bensby:2014}; these errors consist of observational uncertainties and the degeneracies in the stellar parameters. We have achieved the highest precision ever obtained in stellar parameters for this object: $\sigma$ (\teff)=4 K, $\sigma (\log g)=0.012$ dex, $\sigma (v_{t})=0.01 $ km$ \rm{s}^{-1}$, and $\sigma (\feh)=0.004$ dex. \\
In Table \ref{tab:sources} we compare our stellar parameters with previous studies. They all agree within the errors.

\subsection{Age and mass}
\citet{Melendez:2012} and \citet{Nissen:2015} have shown that it is possible to estimate reliable ages of solar-like stars using the isochrone technique \citep[e.g.,][]{Lachaume:1999}. This was made possible by the extremely precise measurements of the stellar parameters and in particular using an accurate log $g$ instead of an uncertain absolute magnitude $\rm{(M_{v})}$ as input parameters. Taking into account the considerations above, \citet{Ramirez:2014-1} developed a python code named \qq (qoyllur-quipu)\footnote{Qoyllur-quipu is a free code based on python, and it is available at \object{https://github.com/astroChasqui/q2.}}, which computes the mass, age, luminosity, and radius. To calculate the age, the \qq\ code uses the isochrone method, adopting a fine grid of Yonsei-Yale isochrones \citep[e.g.,][]{Yi:2001}. Our results are listed in Table \ref{tab:results}. The age we determine (2.0 $\pm$ 0.4 Gyr)\footnote{The probability distribution functions indicate another solution with age younger than 1 Gyr, but this solution was discarded because all other age indicators favor the solution around 2 Gyr.} is also consistent with the high-level activity of this star, which has an average activity indicator of $\log R^{'}_{HK}$ = -4.74  ($\sigma$ = 0.11), according to \citet{Wright:2004}, \citet{White:2007}, and \citet{Isaacson:2010}. We used seven individual $\log R^{'}_{HK}$ measurements available in the references above and the color (B-V) = 0.651 \citep{Ramirez:2012} to determine chromospheric ages from the chromospheric activity indicator $\log R^{'}_{HK}$. Following the relations given in \citet{Wright:2004} and \citet{Mamajek:2008}, we find mean ages of 2.3 $\pm$ 0.4 Gyr and 2.6 $\pm$ 0.6 Gyr, respectively, and a weighted average of 2.4 $\pm$ 0.3 Gyr. This value agrees well with our isochronal age.
\begin{table}
\centering
\captionsetup{justification=centering}
\caption{ Fundamental parameters}
\label{tab:results}
\begin{tabular}{cc}
\hline
\hline
 HIP 100963       & Parameters             \\ \hline
Mass             & $1.03^{+0.02}_{-0.01}$ $\ \mathrm{M_{\odot}}$ \\
Distance         & $28.2 \pm 0.5 $ pc           \\
Convective Mass  & $0.018$ $\ \mathrm{M_{\odot}}$   \\
Age              & 2.0 $\pm$ 0.4 Gyr    \\ 
logL             & $-0.02^{+0.01}_{-0.01} \ \mathrm{L_{\odot}}$      \\
$\mathrm{M_{V}}$ & $4.83^{+0.04}_{-0.03}$ mag            \\
Radius           & $0.97^{+0.01}_{-0.02} \ \mathrm{R_{\odot}}$    \\
\hline \hline
\end{tabular}
\tablefoot{Mass, age, luminosity, and radius were estimated using the \qq \ code, whereas the convective mass was determined by using our code and applying the grid by \citet{Siess:2000}. The $\mathrm{M_{V}}$ and distance were computed using the Hipparcos parallax given by \citet{Van:2007}.}
\end{table}

\subsection{Abundance trends}
We measured high-precision chemical abundances of 26 elements other than Li using EWs: C, O, Na, Mg, Al, Si, Ca, Sc, Ti, V, Cr, Mn, Fe, Co, Ni, Cu, Zn, Sr, Y, Zr, Ba, La, Ce, Nd, Sm, and Eu relative to the Sun. We adopted the hyperfine structure data used by \citet{MelendezIvanAmanda:2014} for V, Mn, Co, Cu, Y, Ba, La, and Eu, with isotopic fractions as given in \citet{Asplund:2009}, \citet{McWilliam:1998}, and \citet{Cohen:2003} for Cu, Ba, and Eu, respectively.

Non-local thermal equilibrium (NLTE) effects in solar twins are negligible \citep{Melendez:2012, Monroe:2013}, but might be important for the O I triplet. Therefore we applied differential NLTE corrections for O using the NLTE grid of \citet{Ramirez:2007}. Recent oxygen NLTE calculations \citep{Amarsi:2015} predict quite different absolute NLTE corrections than those from \citet{Ramirez:2007}, but for our work the effect is negligible because all the calculations are strictly differential. Adopting the grid of \citet{Amarsi:2015} would result in a differential abundance lower by only 0.004 dex.

\begin{figure}
\centering
\includegraphics[width=9cm]{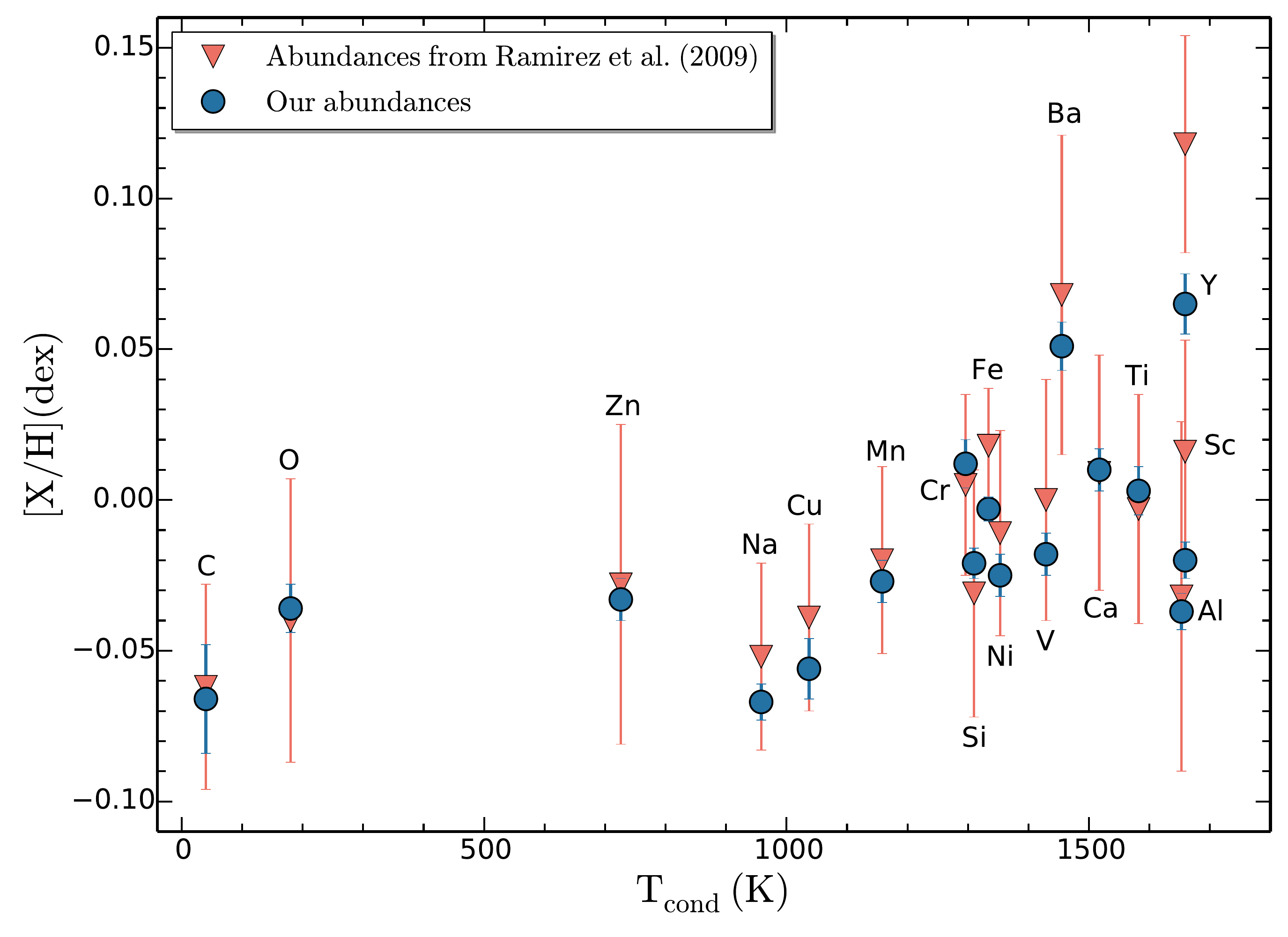}
\caption{Comparison between our abundance pattern and that by \citet{Ramirez:2009} as a function of condensation temperature, T$\rm{_{cond}}$ \citep{Lodders:2003}.}
\label{fig:comparison}
\end{figure}

In Table \ref{tab:adundances} we show our differential abundances relative to the Sun, including observational, systematic, and the total error (obtained from quadratically adding the statistical and systematic errors). Our results agree with the abundances calculated by \citet{Ramirez:2009}, as shown in Fig. \ref{fig:comparison},
but our results are more precise because of the better quality of the data. The mean difference is $\Delta$ [X/H] (\citet{Ramirez:2009} - this work) = 0.011 ($\sigma$ = 0.016 dex).

\citet{Nissen:2015} conducted a similar high-precision chemical abundance analysis of 21 solar twins to investigate abundance trends with age and dust condensation temperature. He found that the [Y/Mg] ratio was correlated with age, so that this abundance ratio offers an independent age indicator. Using his relation, we estimate the age of HIP 100963 as $2.1 \pm 0.4$ Gyr, in agreement with the isochronal age computed by using \qq\ and with the age estimated from a chromospheric diagnostics.

\begin{table*}
\caption{Stellar abundances [X/H] of HIP 100963 relative to the Sun and corresponding errors.}
\label{tab:adundances}
\centering 
\renewcommand{\footnoterule}{}  % to avoid a line before footnotes
\begin{tabular}{lrrrrrrrrrrr} 
\hline    
\hline 
{Element}& [X/H]   & $\Delta \tef$ & $\Delta$log $g$ & $\Delta v_t$ & $\Delta$[Fe/H] & param\tablefootmark{a} & obs\tablefootmark{b} & total\tablefootmark{c} & [X/H]\tablefootmark{d} \\
{}       &   LTE   & $\pm$ 4K      & $\pm$ 0.01 dex& $\pm$ 0.01 km s$^{-1}$  & $\pm$ 0.004 dex   &  &  &  & GCE \\
{}       &   (dex) & (dex)         & (dex)         & (dex)           & (dex)       & (dex)        & (dex) & (dex) & (dex) \\
\hline
C        & -0.066  &  0.002 &  0.003 &  0.000 &  0.000 & 0.004 & 0.018 & 0.018  &  0.015      \\  
O        & -0.036  &  0.004 &  0.002 &  0.001 &  0.000 & 0.005 & 0.006 & 0.008  &  0.038      \\ 
Na       & -0.067  &  0.002 &  0.001 &  0.000 &  0.000 & 0.002 & 0.006 & 0.006  &  0.008     \\  
Mg       & -0.024  &  0.004 &  0.001 &  0.002 &  0.001 & 0.005 & 0.007 & 0.008  &  0.022     \\
Al       & -0.037  &  0.002 &  0.000 &  0.000 &  0.000 & 0.002 & 0.006 & 0.006  & -0.006    \\
Si       & -0.021  &  0.001 &  0.001 &  0.000 &  0.001 & 0.002 & 0.005 & 0.005  &  0.004     \\  
Ca       &  0.010  &  0.003 &  0.001 &  0.002 &  0.001 & 0.004 & 0.006 & 0.007  & -0.009   \\  
Sc*      & -0.020  &  0.000 &  0.004 &  0.002 &  0.001 & 0.005 & 0.004 & 0.006  &  0.048     \\  
Ti*      &  0.003  &  0.000 &  0.005 &  0.002 &  0.001 & 0.005 & 0.006 & 0.008  &  0.028     \\  
V        & -0.018  &  0.004 &  0.001 &  0.000 &  0.000 & 0.004 & 0.006 & 0.007  &  0.002     \\  
Cr*      &  0.012  &  0.001 &  0.004 &  0.002 &  0.001 & 0.005 & 0.006 & 0.008  &  0.002    \\  
Mn       & -0.027  &  0.003 &  0.000 &  0.001 &  0.001 & 0.003 & 0.006 & 0.007  &  0.021     \\  
Fe*      & -0.003  &  0.003 &  0.000 &  0.002 &  0.001 & 0.004 & 0.001 & 0.004  & -0.003   \\  
Co       & -0.046  &  0.003 &  0.002 &  0.000 &  0.000 & 0.004 & 0.011 & 0.012  &  0.017     \\  
Ni       & -0.025  &  0.002 &  0.000 &  0.002 &  0.001 & 0.003 & 0.006 & 0.007  &  0.024     \\  
Cu       & -0.056  &  0.002 &  0.001 &  0.002 &  0.001 & 0.003 & 0.009 & 0.010  &  0.027     \\  
Zn       & -0.033  &  0.000 &  0.001 &  0.002 &  0.001 & 0.002 & 0.007 & 0.007  &  0.033     \\
Sr       &  0.055  &  0.004 &  0.001 &  0.001 &  0.001 & 0.004 & 0.007 & 0.008  &  \ldots \\
Y        &  0.065  &  0.001 &  0.005 &  0.004 &  0.002 & 0.007 & 0.008 & 0.010  &  \ldots \\
Zr       &  0.074  &  0.001 &  0.005 &  0.002 &  0.001 & 0.006 & 0.006 & 0.008  &  \ldots \\
Ba       &  0.051  &  0.001 &  0.001 &  0.004 &  0.002 & 0.005 & 0.008 & 0.009  &  \ldots \\ 
La       &  0.072  &  0.001 &  0.006 &  0.000 &  0.001 & 0.006 & 0.011 & 0.013  &  \ldots \\ 
Ce       &  0.054  &  0.001 &  0.005 &  0.001 &  0.002 & 0.006 & 0.007 & 0.009  &  \ldots \\ 
Nd       &  0.066  &  0.001 &  0.006 &  0.000 &  0.002 & 0.006 & 0.006 & 0.009  &  \ldots \\ 
Sm       &  0.086  &  0.001 &  0.005 &  0.001 &  0.001 & 0.005 & 0.007 & 0.009  &  \ldots \\ 
Eu       &  0.069  &  0.001 &  0.006 &  0.001 &  0.001 & 0.006 & 0.007 & 0.009  &  \ldots \\

\hline       
\end{tabular}
\tablefoot{\\
\tablefoottext{a}{Adding errors in stellar parameters} \\
\tablefoottext{b}{Observational errors} \\
\tablefoottext{c}{Total errors (stellar and observational
parameters)}\\
\tablefoottext{d}{Abundances corrected by subtracting the chemical evolution of the Galactic thin disk}\\
\tablefoottext{*}{The errors due to uncertainties in the stellar parameters come from ScII, TiII, CrII, and FeI.}
}
\end{table*}

\section{Discussion}
\subsection{Light elements (Z $\leq$ 30)}
\begin{figure}
\centering
\includegraphics[width=9cm]{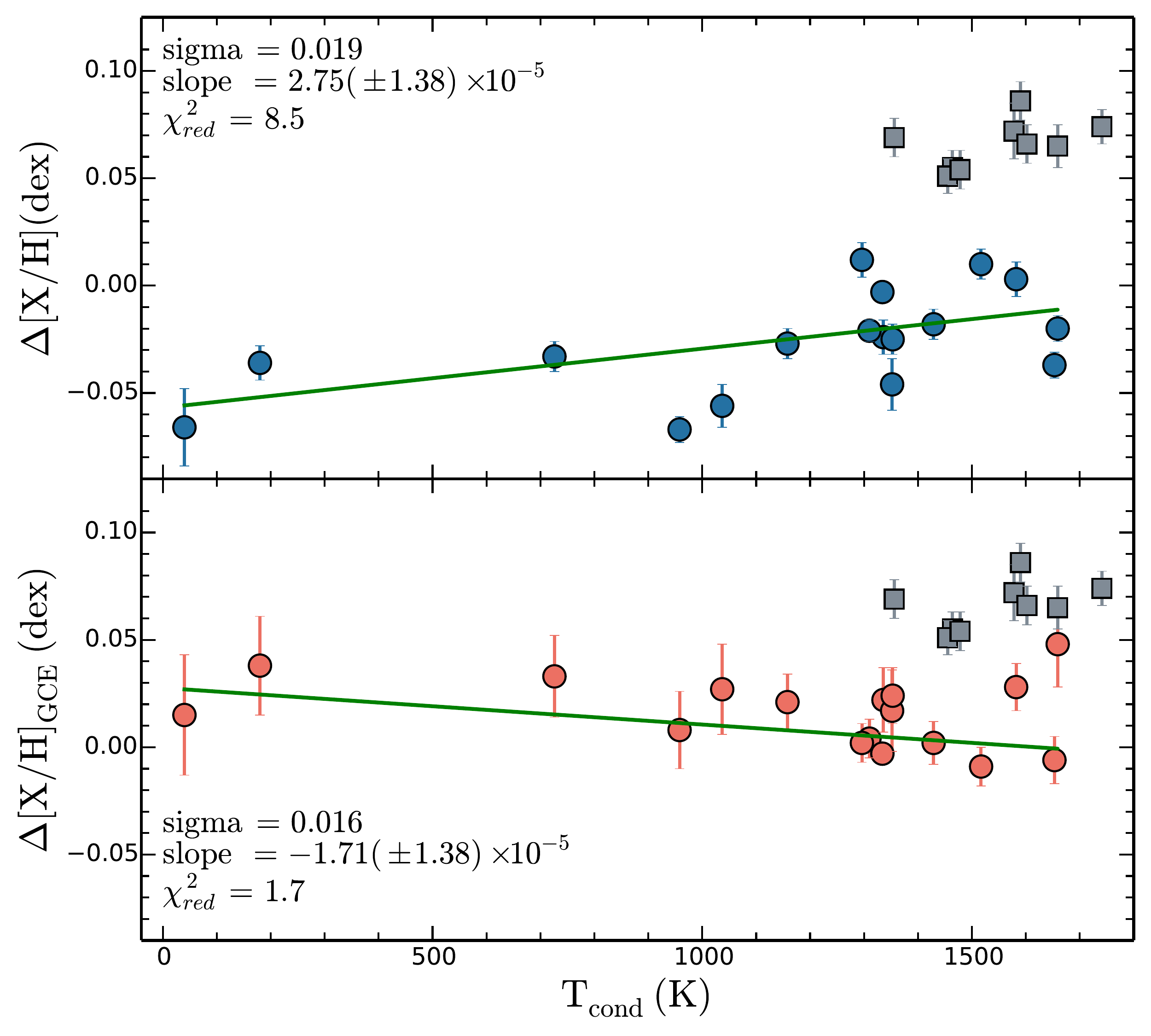}
\caption{Upper panel: Observed differential abundances relative to the Sun versus condensation temperature. Lower panel: Abundance trend after correcting for the GCE to the solar age. Linear fits to the data (excluding the neutron-capture elements) are overplotted in each panel as a green solid line. Neutron-capture elements are represented by gray filled squares. Sigma represents the dispersion about the linear fit.}
\label{fig:fit-ligher-elements}
\end{figure}

The abundance pattern of HIP 100963 relative to the Sun is shown in Fig. \ref{fig:fit-ligher-elements} (blue filled circles in the upper panel). The fit of their light elements (Z $\leq$ 30) gives a positive slope with an average dispersion of 0.019 dex, which is higher than the average error bar (0.008 dex). This might be due to the scatter introduced by the GCE. 

A correlation between T$_{\rm{cond}}$ trend and age was first observed by \citet{Ramirez:2014-1}.  \cite{Adibekyan:2014A} also
showed a correlation between the abundance trends versus condensation temperature and stellar ages. They analyzed a sample of 148 solar-like stars and observed that young stars are richer in refractories than old stars. They suggested that this behavior is due to the chemical evolution of the Galactic thin disk. Later, \citet{Nissen:2015} and \citet{Spina:2015} confirmed this correlation and concluded that the $T\rm{_{cond}}$-slope decreases with increasing stellar age. 


We employed the [X/Fe]-age relation of \citet{Spina:2015} to correct for the GCE for elements with Z $\leq$ 30. As HIP 100963 is younger than the Sun, we added the effect of the GCE to its elemental abundances because our differential abundances are relative to the present-day Sun. This yields
\begin{equation}
\mathrm{[X/H]_{GCE}} = \mathrm{[X/H]} + b \times (\rm{Age_{Sun}} - \rm{Age_{twin}}),
\end{equation}
where $b$ is the slope found by \citet{Spina:2015}\footnote{\citet{Spina:2015} performed a linear fit ([X/Fe] = age$\times$ b + a) with their parameters listed in Table 3 of their work. The fits are used to correct for the GCE as a function of age, but do not take possible inhomogeneous GCE effects into account.}. In Fig. \ref{fig:fit-ligher-elements} (lower panel) we show the elemental abundance differences ($\mathrm{[X/H]}_{GCE}$) \ adding the GCE effect between HIP 100963 and the Sun (red filled circles) as a function of condensation temperature \citep{Lodders:2003}.
We did not correct for the n-capture elements (gray squares) because \citet{Spina:2015} did not include most of these elements. The average dispersion from the fit is 0.016 dex, in agreement with the average error bar (0.015 dex), which now includes the uncertainty in the GCE correction (also including the error in age). Besides that, the $\chi^{2}_{red}$ decrease to 1.7 from the initial value of 8.5. Within the errors the abundance pattern of HIP 100963 corrected by GCE (to the solar age) is similar to the Sun, or perhaps slightly more depleted in refractories than the Sun. Notice that the slope changed from 2.8 to -1.7, and that the difference is significant at the $\sim$ 2 sigma level, i.e., the contribution of GCE is important.

\citet{Melendez:2009} were the first to find that the Sun is deficient in refractories when compared to the average abundance of 11 solar twins. They argued that the peculiar solar abundance pattern is a mirror image of the abundance pattern of meteorites and inner rocky planets in our solar system. This suggests that the refractory-poor solar pattern may be linked to the formation of terrestrial planets in the solar system. If the \citet{Melendez:2009} scenario is correct, the solar pattern of HIP 100963 means that it may have formed rocky planets. 

HIP 100963 is included in the California Planet Search \citep{Isaacson:2010}, but no planet has been detected yet. Only a few data points have been reported so far, however.

\subsection{Neutron capture elements (Z > 30)}
\begin{figure}
\centering
\includegraphics[width=9cm]{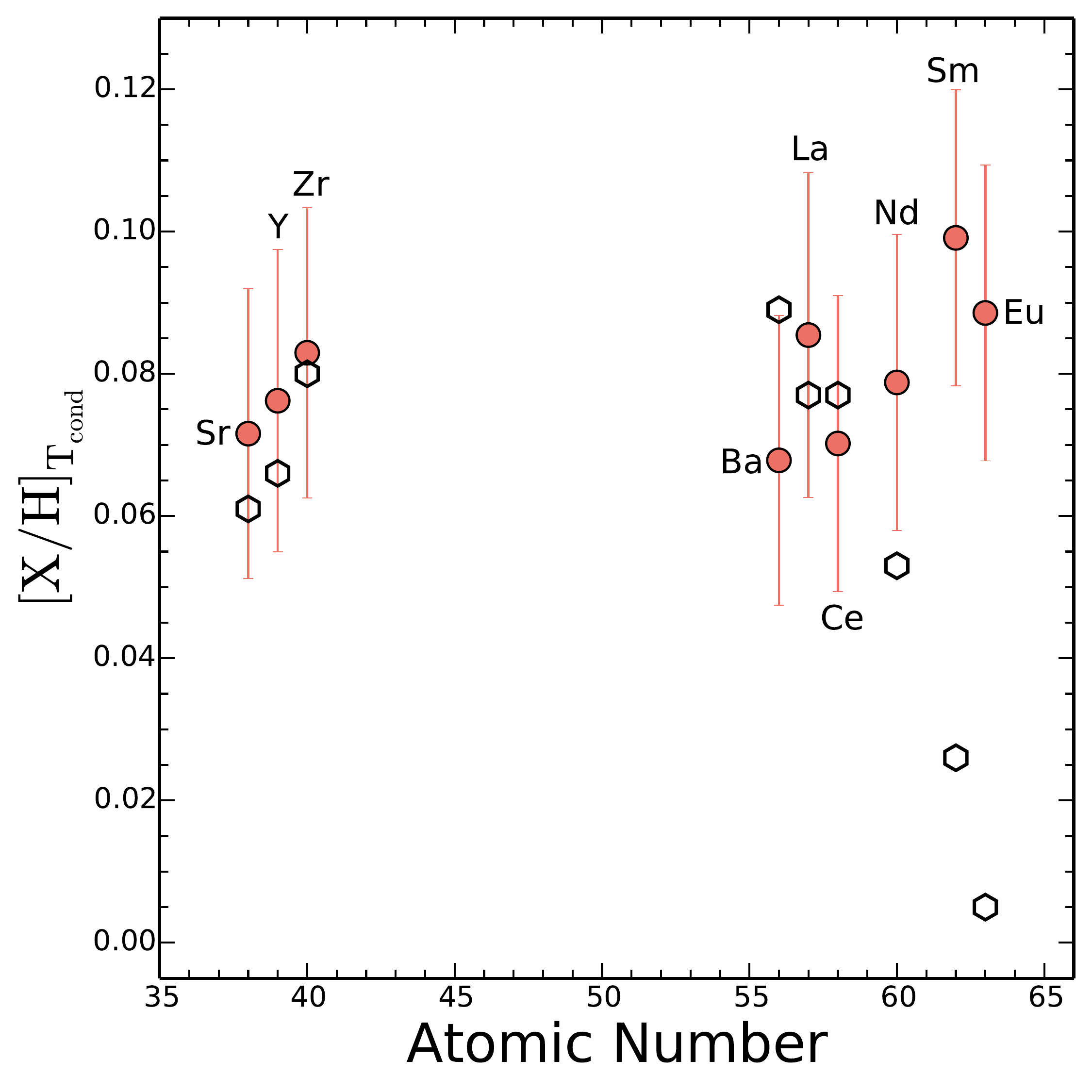}
\caption{Comparison of the $\rm{[X/H]_{T_{cond}}}$ (red filled circles) ratios with predicted abundances from the AGB pollution (open hexagons).} 
\label{fig:Xt-Z}
\end{figure}

As can be seen in Fig. \ref{fig:fit-ligher-elements}, HIP 100963 presents significant enhancements in neutron-capture elements (gray squares) from the $s$- and $r$-process. For the solar twin 18 Sco, \citet{MelendezIvanAmanda:2014} argued that these enhancements might be due to pollution from asymptotic giant branch (AGB) stars and $r$-process sources, which enrich the interstellar medium over time. Although massive stars dominate the production of the lighter $s$-process elements (Sr, Y, and Zr) at low metallicites, at solar metallicity the production of these elements is dominated by AGB stars \citep{Bisterzo:2014}. We here followed the procedure adopted by \citet{MelendezIvanAmanda:2014} for the solar twin 18 Sco. Assuming that the observed enhancements are due to the pollution from AGB stars, we fit the lighter elements $\rm{[X/H]_{Z \leq 30}}$  to subtract the trend with condensation temperature from heavy elements $\rm{[X/H]_{Z>30}}$,
\begin{equation}
\rm{[X/H]_{Tcond}} = \rm{[X/H]_{Z>30}} - (-0.056 + 2.753\times 10^{-5} \rm{T_{cond})}, 
\end{equation}
where $\rm{[X/H]_{Tcond}}$ represents the abundance of neutron-capture elements corrected for the condensation temperature effect, whereas the last two terms come from the fit of the lighter abundances trend versus condensation temperature without GCE (see Fig. \ref{fig:fit-ligher-elements}). To compute the amount of pollution of the protocloud by an AGB star ($\rm{[X/H]_{AGB}}$), we used the same model as adopted by \citet{MelendezIvanAmanda:2014}\footnote{This is an AGB stellar model with 3 \sm\ as initial mass, and Z = 0.01 of metallicity \citep{Karakas:2010, MelendezIvanAmanda:2014}.}, with diluted yields of a small fraction of AGB ejecta into a protocloud of
solar composition of 1 \sm \  \citep{Asplund:2009}. We estimate a dilution of 0.34\% mass of AGB material to match the average observed enhancements in the $s$-process elements (Sr, Y, Zr, Ba, La, and Ce). The two abundances $\rm{[X/H]_{Tcond}}$ and $\rm{[X/H]_{AGB}}$ are given in Table \ref{tab:neutron-capture}. Figure \ref{fig:Xt-Z} shows that the $s$-process elements Y, Ba, La, and Ce are well reproduced, but the $r$-process elements (Sm and Eu) are not. Therefore, the AGB contribution alone cannot explain the heavy abundance enhancement. Another source of enhancement is the contribution from the $r$-process. To determine this enrichment, we first subtracted the AGB contribution from the heavy elements,
\begin{equation}
\rm{[X/H]}_{\rm{r-process}} = \rm{[X/H]_{Tcond}} - \rm{[X/H]_{AGB}}.
\end{equation}
Then we compared these ratios with the predicted enhancements based on the $r$-process in the solar system \citep{Simmerer:2004, Bisterzo:2014}. \citet{MelendezIvanAmanda:2014} determined the $r$-process contribution, which is given by 
\begin{equation}
\rm{[X/H]_{SS-r}} = \rm{\Delta \tau} \log_{10} r_{SS} + \rm{[X/H]_{Tcond}},
\end{equation}

\begin{figure}
\centering
\includegraphics[width=9cm]{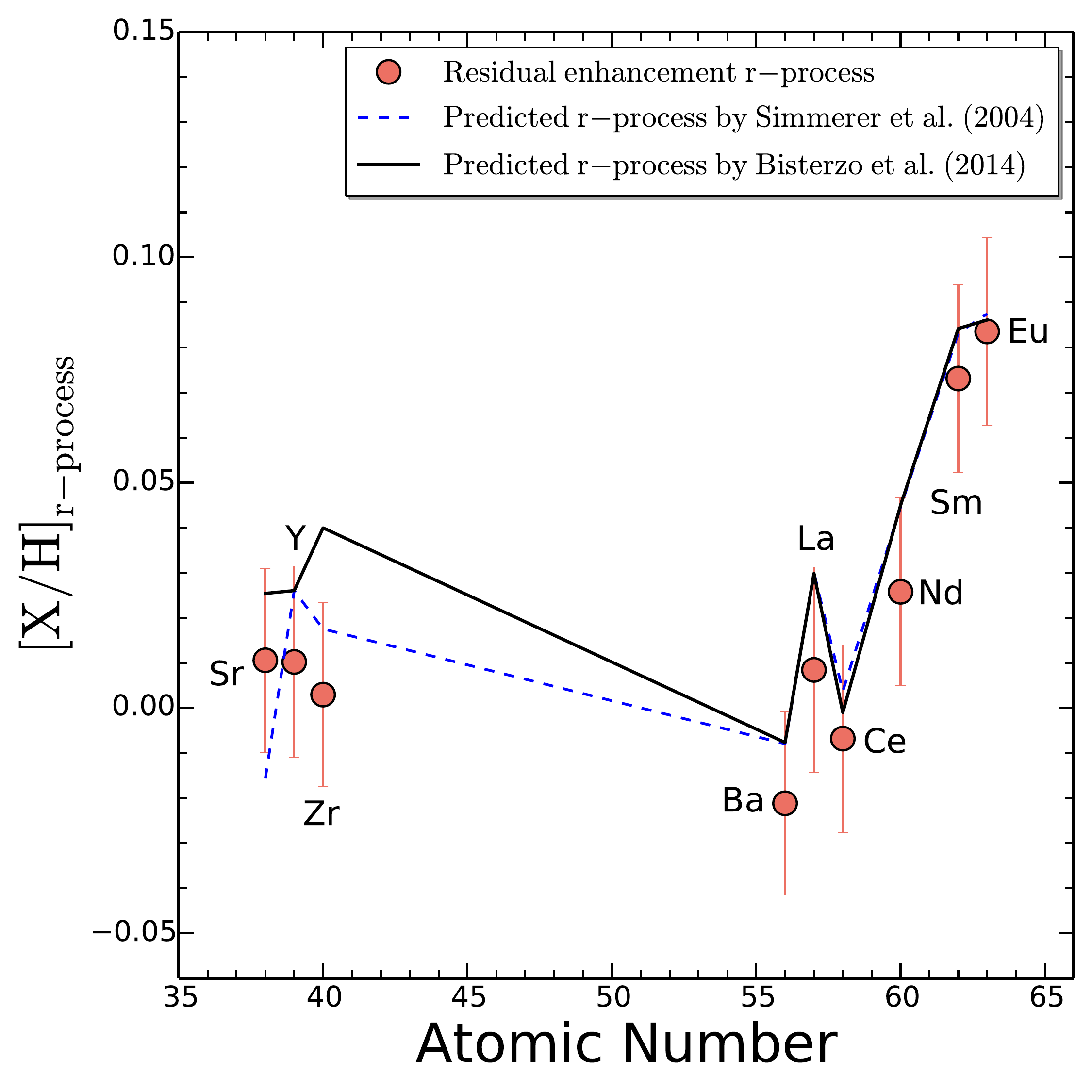}
\caption{Residual enhancement $\rm{[X/H]_{r-process}}$ due to the contribution of the $r$-process. Predicted enhancement by \citet{Bisterzo:2014} and \citet{Simmerer:2004} are also plotted as solid and dashed lines, respectively.}
\label{fig:Xt-Xr-Z}
\end{figure}

where $\Delta \tau$ is the average of the two most enhanced $s$-process and $r$-process elements ($\Delta \tau$ = 0.091 dex), and $\rm{r_{SS}}$ is the $r$-process fraction in the solar system. In Fig. \ref{fig:Xt-Xr-Z}
we show our results, which  agree reasonably well with the predicted $r$-process in the solar system \citep{Simmerer:2004, Bisterzo:2014}. Even for elements that are partially produced through the contribution of $r$-process (e.g., Nd), we obtain agreement within the error. Our results, together with that obtained for 18 Sco \citep{MelendezIvanAmanda:2014}, suggest an enhanced contribution of $r$- and $s$-process elements in young stars. Our results are also supported by \citet{Dorazi:2009}, \citet{Maiorca:2011}, and \citet{Dorazi:2012}, who showed that younger objects exhibit $s$-process enhancements in open clusters.

\begin{table}
\caption{Enhancement of neutron-capture elemental abundances and their decomposition in HIP 100963}
\label{tab:neutron-capture}
\centering 
\renewcommand{\footnoterule}{}  % to avoid a line before footnotes
\begin{tabular}{lrrrr} 
\hline 
\hline 
{Z}      & {Element}   & $[X/H]_{Tcond}$ & $[X/H]_{AGB}$ &  $[X/H]_{r -  process}$ \\
{}       &             &   (dex)     &        (dex)  &           (dex) \\
\hline
38 &  Sr &  0.071 &  0.061 &  0.010 \\
39 &  Y  &  0.076 &  0.066 &  0.010 \\
40 &  Zr &  0.083 &  0.080 &  0.003 \\
56 &  Ba &  0.067 &  0.089 & -0.020 \\ 
57 &  La &  0.085 &  0.077 &  0.008 \\ 
58 &  Ce &  0.070 &  0.077 & -0.007 \\ 
60 &  Nd &  0.078 &  0.053 &  0.025 \\ 
62 &  Sm &  0.099 &  0.026 &  0.073 \\ 
63 &  Eu &  0.088 &  0.005 &  0.083 \\

\hline       
\end{tabular}
%\tablefoot{.}
\end{table}

\subsection{Lithium abundance}
\label{sec:4}
Previous works suggest a strong connection between lithium abundance and stellar age \citep{Soderblom:1983, Charbonnel:2005, Denissenkov:2010, Baumann:2010, Monroe:2013, MelendezSchirbel:2014} based on findings  that showed younger stars to have large Li abundances. To determine
whether HIP 10093 follows this relation, we estimated the lithium abundance following the procedure explained in \citet{Melendez:2012}, which consists of first estimating the line broadening and then performing spectral synthesis of the Li feature. We computed the stellar macroturbulence velocity \citep{Tucci:2015} through the relation
\begin{equation}
\label{eq:macro}
v_{macro} = v_{macro,\ \odot} + (\rm{T_{eff}} - 5777)/486.
\end{equation}
We fixed \vsini$_{\ \odot}$ = 1.9 km$\ \rm{s}^{-1}$ \citep{Bruning:1984, Saar:1997} and obtained by spectral synthesis of six lines (see below) $v_{macro,\ \odot}$ = 3.34 km$\ \rm{s}^{-1}$ , which was employed in Eq. (\ref{eq:macro}) to estimate the $v_{macro}$ of HIP 100963, which in turn  yielded = 3.43 km$ \rm{s}^{-1}$. Then the projected rotational velocity (\vsini) was measured by fitting the line profiles of five iron lines: $6027.050\ \AA, 6093.644\ \AA, 6151.618\ \AA, 6165.360\ \AA, 6705.102\ \AA,$ and one Ni I line at $6767.772\ \AA$. We found \vsini\ = 1.93 $\pm$ 0.21 km\ $\rm{s}^{-1}$, similar to what was determined by \citet{Takeda:2009} (\vsini\ = 2.04 km\ $\rm{s}^{-1}$; no error bar was given), and in good agreement with the value 1.9$\ \pm\ $0.3 km$\rm{s}^{-1}$ of \citet{Notsu-3:2015} \footnote{The value found by \citet{Notsu-3:2015} for HIP 100963 is 2.4 $\pm$ 0.3 km $\rm{s}^{-1}$, with \vsini$_{\odot}$ = 2.4 km $\rm{s}^{-1}$. Thus, the result relative to the Sun is $\Delta$ \vsini = 0.0 km  $\rm{s}^{-1}$. The same is true for \citet{Takeda:2009}, where \vsini$_{\odot}$ = 2.13 km $\rm{s}^{-1}$ and \vsini = 2.27 km $\rm{s}^{-1}$ for HIP 100963.}.

\begin{figure}
\centering
\includegraphics[width=9cm]{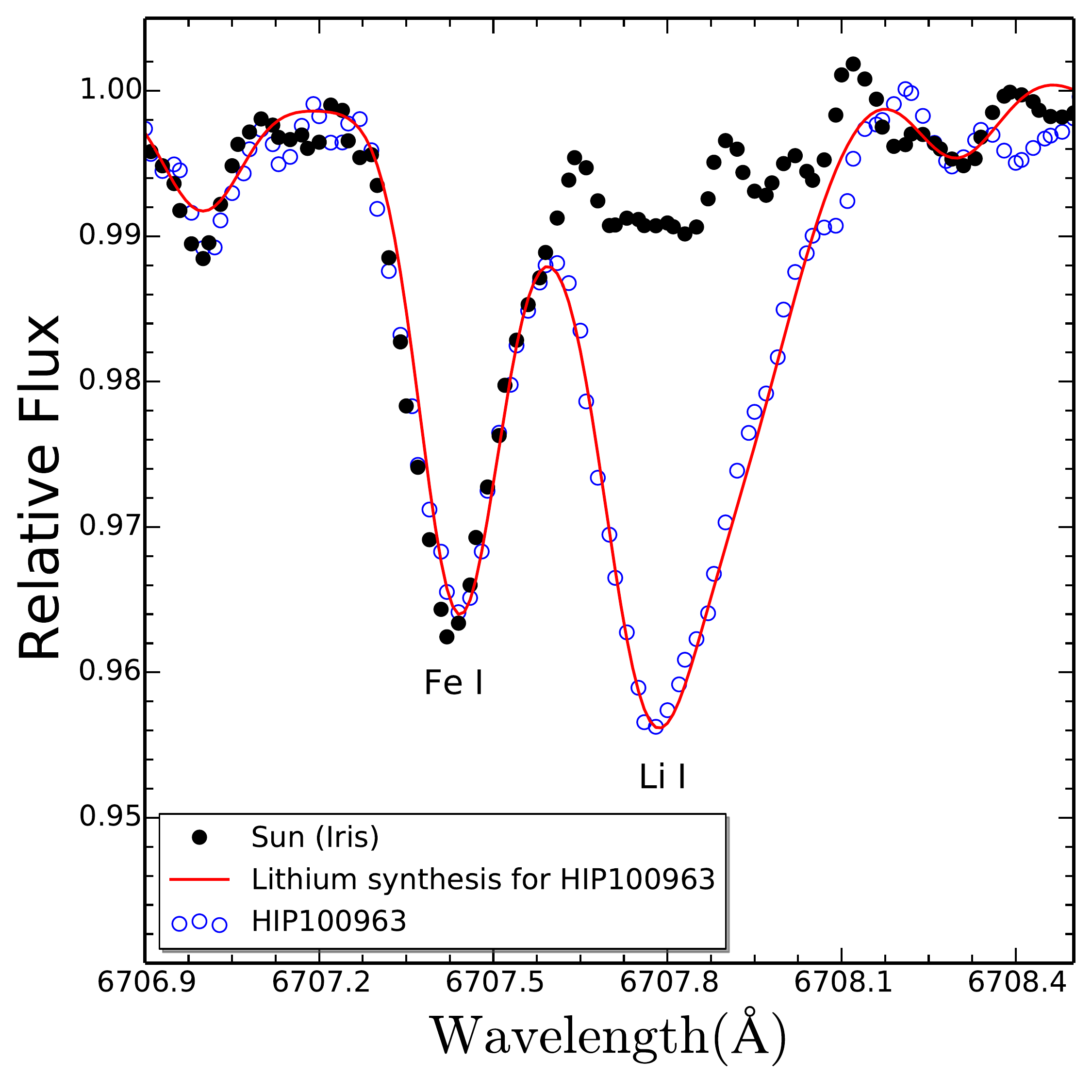}
\caption{Spectral synthesis of the Li doublet at 6708 A. Open and filled circles represent the spectra of HIP 100963 and the Sun, respectively, while the synthetic spectrum is overplotted as a red solid line.}
\label{fig:Lythium-spectra}
\end{figure}

\begin{figure}
\centering
\includegraphics[width=9cm]{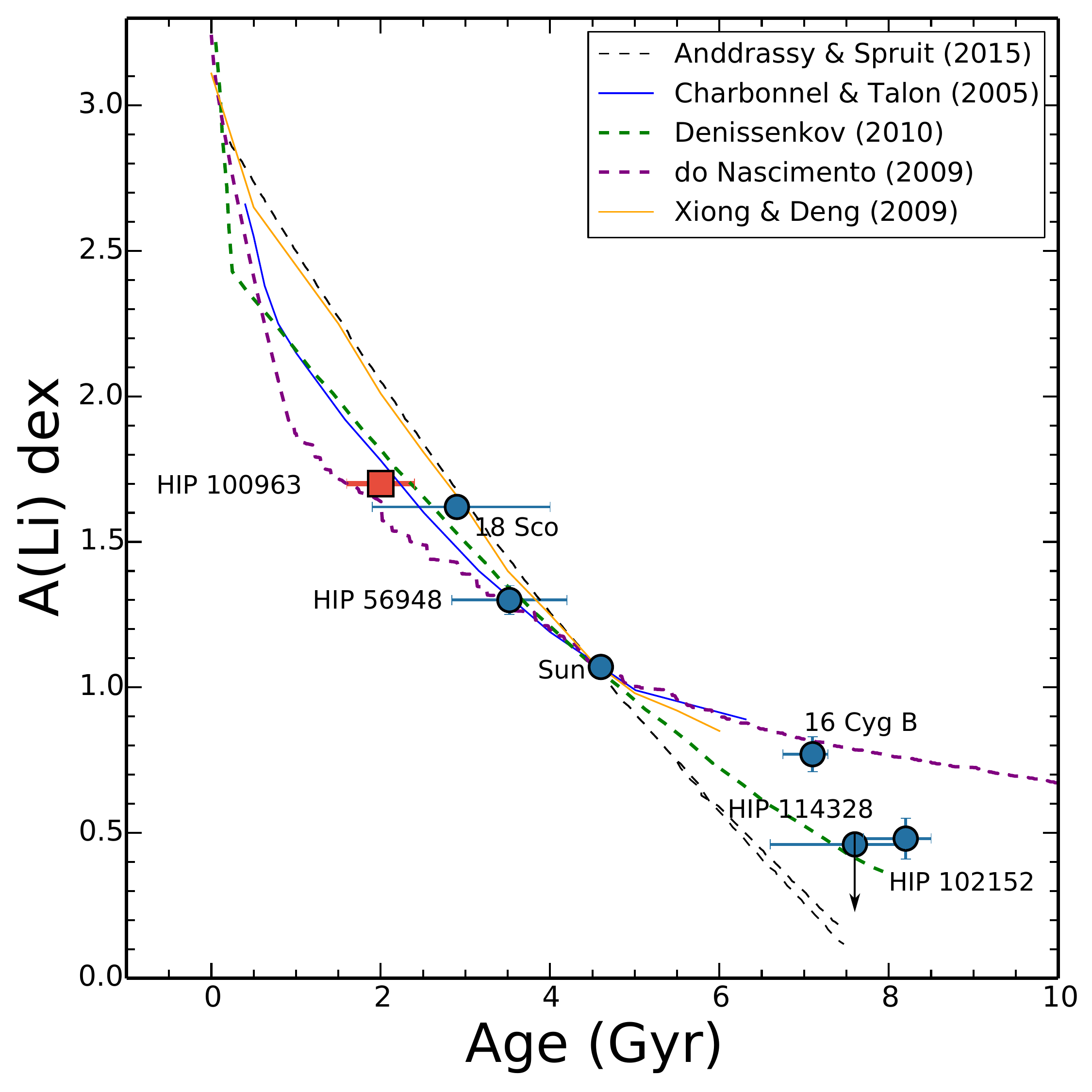}
\caption{Li vs. age for HIP 100963 (red filled square) based on NLTE analysis and isochronal age. The Li abundance for other stars analyzed by our group in previous works \citep{Ramirez:2011, Melendez:2012, Monroe:2013, MelendezIvanAmanda:2014} is shown with blue filled circles. The lithium abundance errors are smaller than the size of the circles. Li depletion models are also shown with different lines.} %The age of HIP 114328 was update by \citet{Marilia:2015}.
\label{fig:Andrassy-Li-Age}
\end{figure}

 HIP 100963 (blue open circles) presents a deeper Li feature
in Fig. 7 than that in the Sun (black filled circles). Based
on the $v_{macro}$ and \vsini\ we determined above, we performed the spectral synthesis of the Li line adopting the line list of \citet{Melendez:2012} and using the $synth$ driver in MOOG.
This yielded an LTE abundance of 1.67 dex. The lithium abundance in NLTE is $1.70 \pm 0.03$; it was obtained by using the NLTE corrections of \citet{Lind:2009}\footnote{INSPECT database available at \object{wwww.inspect-stars.com}}. The A(Li) uncertainties were obtained taking into consideration errors due to S/N, continuum setting, and stellar parameters. The quadratic sum of the three uncertainty sources gives an error of 0.03 dex. Our result closely agrees with the Li abundance\footnote{No error bars were given by \citet{Takeda:2007} and \citet{Notsu-3:2015}.} of 1.72 dex and 1.7 dex determined by \citet{Takeda:2007} and \citet{Notsu-3:2015}, respectively.  

\citet{Melendez:2012} analyzed the Li-age correlation using the Sun and two solar twins (18 Sco and HIP 56948). They concluded that lithium decays with age. Additional works \citep{Ramirez:2011, Monroe:2013, MelendezSchirbel:2014} confirmed their results \citep[see also][]{Baumann:2010}. Figure \ref{fig:Andrassy-Li-Age} shows
that HIP 100963 follows the Li-age correlation of previous works, extending the result to stellar ages younger than the Sun. We also compared our findings with non-standard Li depletion models developed for the Sun \citep{Charbonnel:2005, Nascimento:2009, Xiong:2009, Denissenkov:2010, Andrassy:2015}. As can be seen, the agreement is very good. These previous models include transport mechanisms that are not considered in the standard solar model.

\section{Conclusions}
\label{sec:5}
We have performed a high-precision differential study of the solar twin candidate HIP 100963 based on Keck+HIRES spectra. The atmospheric parameters that we found are $T_{\rm{eff}}=5818 \pm 4$ K, log $g = 4.49 \pm 0.01$ dex, $v_{t} = 1.03 \pm 0.01 $ km$\ \rm{s}^{-1}$ , and [Fe/H] $ = -0.003 \pm 0.004$ dex. While these values agree with previous studies, we have achieved considerably higher accuracy and  confirm that HIP 100963 is a solar twin.

We obtained a precise mass and age ($1.03^{+0.02}_{-0.01}$ \sm, 2.0 $\pm$ 0.4 Gyr) that both agree with the values reported in the literature. The abundance trend with T$_{\rm{cond}}$ before the GCE correction is comparable with the enhancement of refractories seen in other solar twins, but considering corrections for the GCE to the same age as the Sun, the abundance pattern is about solar. This means that this star may have formed rocky planets. The abundances of the $s$- and $r$-process elements are enhanced relative to the Sun. This is most likely because HIP 100963 is younger than the Sun.

Finally, the lithium abundance of HIP 100963 falls within the correlation between Li and age suggested in the literature, extending the age coverage that can be used
to test models of Li depletion to relatively young ages.

\begin{acknowledgements}
J.Y.G. acknowledges support by CNPq. J.M thanks for support from FAPESP (2012/24392-2). We are grateful to the many people who have worked to make the Keck telescope and its instruments a reality and to operate and maintain the Keck Observatory. The authors wish to extend special thanks to those of Hawaian ancestry on whose sacred mountain we are privileged to be guests. Without their generous hospitality, none of the observations presented herein would have been possible. D.Y., F.L., M.A. and A.K. gratefully acknowledge support from the Australian
Research Council (grants DP120100991, FL110100012, FT140100554, DP150100250 and FT110100475).
\end{acknowledgements}

\bibliographystyle{aa}
\bibliography{Objects.bib}

\begin{appendix}

\end{appendix}

\end{document}